\documentclass[12pt,preprint]{emulateapj}
\usepackage{amsmath, natbib, placeins, epsfig}
\begin{document}

%\setpagewiselinenumbers
%\modulolinenumbers[1]
%\linenumbers

\title{Exploring the Nature of the Galactic Center Gamma-Ray Source\\ with the Cherenkov Telescope Array}
\author{Tim Linden$^{1}$ and Stefano Profumo$^{1,2}$}
\affil{$^1$ Department of Physics, University of California, Santa Cruz, 1156 High Street, Santa Cruz, CA, 95064}
\affil{$^2$ Santa Cruz Institute for Particle Physics, University of California, Santa Cruz, 1156 High Street, Santa Cruz, CA, 95064}
\shortauthors{}

\begin{abstract}
Observations from multiple gamma-ray telescopes have uncovered a high energy gamma-ray source spatially coincident with the Galactic center. Recently, a compelling model for the broad-band gamma-ray emission has been formulated which posits that high energy protons emanating from Sgr A* could produce gamma-rays through $\pi^0$ decays resulting from  inelastic collisions with the traversed interstellar gas in the region. Models of the gas distribution in the Galactic center region imply that the resulting $\gamma$-ray morphology would be observed as a point source with all current telescopes, but that the upcoming Cherenkov Telescope Array (CTA) may be able to detect an extended emission profile with an unmistakable morphology. Here, we critically evaluate this claim, employing a three dimensional gas distribution model and a detailed Monte Carlo simulation, and using the anticipated effective area and angular resolution of CTA. We find that the impressive angular resolution of CTA will be key to test hadronic emission models conclusively against, for example, point source or dark matter annihilation scenarios. We comment on the relevance of this result for searches for dark matter annihilation in the Galactic center region.
\end{abstract}

\section{Introduction}
\label{sec:introduction}

%Hess observations of the galactic center point source
Early observations from the High Energy Spectroscopic System (H.E.S.S) opened a new window into $\gamma$-ray observations of the Galactic Center (GC) \footnote{Throughout this work, we will employ the term Galactic Center (GC) to refer to both the dynamical center of the Milky Way, as well as to the position of the radio source Sgr A*, which we will consider to be equivalent.} region, including the detection of a bright TeV point source localized to within 1' of the GC. The spectrum of the $\gamma$-ray source is fairly hard, following a power-law $\alpha$~=~-2.2 +/- 0.09 (stat) +/- 0.15 (sys) with a high-energy cutoff exceeding 10~TeV~\citep{2004A&A...425L..13A}. Further observations succeeded in localizing the center of the point-source to within 13" of the GC, and found 85\% of the total $\gamma$-ray emission to be confined within 3~pc (1.2') of the GC \citep{acera_hess_gc_position_2010}.

%HESS View of the High Energy Galactic Center
While the observed morphological details strongly suggest that the TeV $\gamma$-ray signal stems from a point-source spatially coincident with the black hole at the GC, the steady-state nature of the H.E.S.S. source indicates the emission may be originating farther from the GC. While lower energy X-ray and radio observations have uncovered significant variability from Sgr A* on timescales stretching from minutes to years, no variability has yet been observed in $\gamma$-ray observations~\citep{hess_no_variability_2009}. Most notably, a simultaneous observation with H.E.S.S. and Chandra found that an X-ray outburst observed in 2007 was not correlated with any change in the $\gamma$-ray emission~\citep{hess_no_variability_xray}. This implies that the source of the $\gamma$-ray emission may be distinct from the source of low-energy photons. Several models have been posited which would naturally explain an intense TeV $\gamma$-ray emission which is uncorrelated with the lower-energy regime, including photons from dark matter annihilation~\citep{hooper_tev_dm_2004, profumo_tev_dm_2005, hess_galactic_center_dm_2006}, as well as $\pi^0$-decay resulting from the emission of high energy protons from the central black hole and their subsequent collisions with Galactic gas~\citep{aharonian_nernov_hadronic_gc_2005, liu_hadronic_gc_2006a, liu_hadronic_gc_2006b, fryer2007, ballantyne_hess_tev}.

%Fermi Data and Chernyakova et al. result
With the launch of the Fermi-LAT in 2008, the window was opened to observe the GeV $\gamma$-ray spectrum with similar angular and energy resolution to that of H.E.S.S.~\citep{2009ApJ...697.1071A}, unveiling a distinctly different spectral shape from the very high-energy regime. Specifically, an excess of 1-10~GeV $\gamma$-rays was uncovered in the GC compared to that expected from an extrapolation of the TeV source spectrum to GeV energies \citep{vitale_galactic_center, hooper_goodenough_gc, hooper_linden_gc, hooper_kelso_gc}. Several models have been postulated to explain this excess emission, including the annihilation of light, leptophilic dark matter particles \citep{hooper_goodenough_gc, hooper_linden_gc} and emission from millisecond pulsars~\citep{abazajian_msp_gc}. Recently, \citet{chernyakova} re-examined an extension of the hadronic scenario described above down to GeV energies and found that the entirety of the GeV-TeV spectrum could be explained by inelastic processes initiated by protons whose spectrum would follow a single power-law. The softening of the $\gamma$-ray signal at energies $\sim$10-100~GeV was then enforced by fine-tuning the diffusion constant in order to produce diffusive propagation at GeV energies and rectilinear propagation at TeV energies.

%Results from our previous morphology paper
Subsequently, \citet{linden_gc} examined the expected morphology of the hadronic emission model described by \citet{chernyakova}, using a realistic model for the morphology of Galactic gas in the inner 10~pc around Sgr A* as determined by \citet{ferriere_2012_gas_density}. They found that the morphology of TeV emission closely matched observations by\citet{hess_galactic_center_dm_2006} signaling that 85\% of the $\gamma$-ray emission from the GC was confined to within $\sim$3~pc of Sgr A*. Additionally, \citet{linden_gc} found that the energy dependence of this morphology is minimal, and thus the majority of GeV emission detectable by the Fermi-LAT should also reside within 3~pc of the central black hole, which may be in tension with an observed extension of the GeV $\gamma$-ray source at the GC \citep{hooper_goodenough_gc, hooper_linden_gc}. We note that disentangling an extended emission in this region is especially problematic, given the poor knowledge of the diffuse Galactic emission \citep{vitale_galactic_center}. Finally, \citet{linden_gc} noted that while the hadronic emission model should appear point-like to all current $\gamma$-ray instruments, the upcoming Cherenkov Telescope Array (CTA), should observe an extended spatial morphology which would distinguish this scenario from other point-source emission mechanism occurring at the position of Sgr A*.  

%What do we do here
In this {\em letter}, we closely investigate several models for the TeV $\gamma$-ray emission at the GC which may be observed by CTA, including a $\gamma$-ray point source at the position of Sgr A*, p-p collisions due to hadronic emission from the position of Sgr A* \citep{chernyakova, linden_gc}, and dark matter \citep{hooper_tev_dm_2004, profumo_tev_dm_2005, hess_galactic_center_dm_2006}. Specifically, we show that the improved angular resolution of CTA can differentiate these scenarios with surprising accuracy based on morphology alone, allowing for the construction of a rigorous model for the $\gamma$-ray emission from the GC center source. In turn, the resulting morphology observed by CTA will stand as a crucial ingredient in the understanding of high-energy emission from the entire GC region, including possibly the extraction of a  dark matter annihilation signal.

\section{Models}
\label{sec:models}

\begin{figure*}
                \plotone{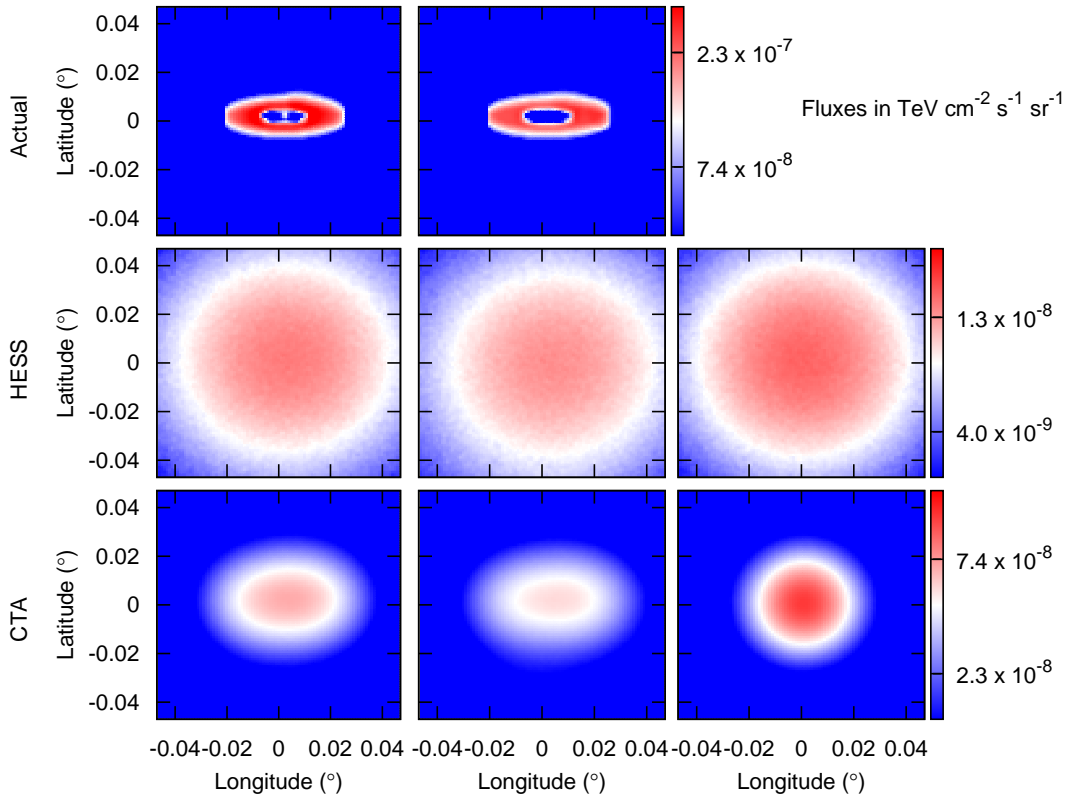}
                	\caption{ \label{fig:2dplot} Observed flux at the solar position as a function of the angle from the galactic center from the actual emission morphology (i.e. for a machine with a perfect angular resolution, top row), the emission as observed by an instrument with the angular resolution of the HESS telescope (middle row) and for an instrument with the angular resolution of the CTA telescope (bottom row) in the case of hadronic emission in the regime rectilinear proton propagation (left column), hadronic emission in the regime of diffusive proton propagation (middle column), and point source emission. The fluxes are shown logarithmically and binned to regions of 10$^{-4}$$^\circ$. The pointsource emission for an instrument with perfect angular resolution is not shown, as it would would provide a delta function at the center of the image. The flux shown stands as the flux between 1-10 TeV in energy averaged over 10000 realizations of 93 hour HESS observations and 1000 realizations of 500 hour CTA observations.}
\end{figure*}

%%Models of the Galactic Center Gas Density (Ferriere 2012)
The morphology of the $\gamma$-ray emission from $\pi^0$ decay is dominated by the distribution of Galactic gas. In order to produce a rigorous model which takes into account the full 3D morphology of the target gas density, we employ the maps of \citet{ferriere_2012_gas_density} which include not only contributions from a spherical diffuse halo, but also from structures associated with the SNR Sgr A East, from belts of molecular clouds, and most importantly, from the high-density gaseous disk known as the circum-nuclear ring. This ring-shaped cloud of gas is located between 1-3~pc from the GC, inclined 20$^\circ$ with respect to the Galactic plane, and contains gas densities approximately two orders of magnitude larger than in the surrounding GC medium~\citep{circumnuclear_ring_1982, 2005ApJ...623..866B}. In this work, we assume the central position and gas density for each gas structure, and set the volume filling factor of each structure to match those provided in Table 1 of \citet{ferriere_2012_gas_density}. Given the high energy of the injected protons in $\gamma$-ray models, we ignore all information on the temperature distribution of Galactic gas, as it is inconsequential for $\gamma$-ray production.

%%Models for Cosmic Ray propagation (Linden et al. 2012)
In their analysis of the combined GeV and TeV spectrum, recent work by both \citet{chernyakova} and \citet{linden_gc} employed a diffusion constant tuned in order to provide a sharp transition between diffusive propagation in the GeV energy regime and rectilinear transport at TeV energies. This transition, which occurs at diffusion constants of approximately 1.4~x~10$^{29}$~cm$^2$s$^{-1}$ for a diffusion zone of 10~pc, must be finely set in order to correctly explain the extremely soft emission spectrum at energies of approximately 100-500~GeV. Thus, a generic prediction of all scenarios which employ a single proton injection spectrum to correctly match both the GeV and TeV $\gamma$-ray emission spectrum, is cosmic ray propagation which is transitioning from the diffusive to the rectilinear regime at TeV energies. In the rectilinear regime, the cosmic-ray density falls as r$^{-2}$, while in the diffusive regime the cosmic-ray density falls instead as r$^{-1}$, due to the square dependence of the escape time on the size of the diffusion region.

In either case, we will assume that cosmic ray protons in this energy range tend to interact with gas less than one time before leaving the diffusion region, validating the assumption that the energy spectrum of cosmic ray protons is position independent. For the gas density maps provided by \citet{ferriere_2012_gas_density} this assumption holds so long as the average value of the diffusion constant in the inner 10~pc of the galaxy exceeds 1.8~$\times$~10$^{27}$~cm$^2$~s$^{-1}$.  Under this restriction, coupled with the assumption that the $\gamma$-ray emission from the galactic center is in steady state, our models for the morphology of the $\gamma$-ray emission depend only on the radial density of cosmic-ray protons, rather than the exact diffusion scenario which produces the proton morphology.

Throughout this work, we will calculate the expected $\gamma$-ray morphology for both the diffusive and rectilinear models of proton-propagation, finding that the qualitative arguments presented throughout this paper do not depend on the exact diffusion model considered. Furthermore, we can consider these two scenarios to bound the distributions expected for relativistic particle motion due to interactions with magneto-hydrodynamic waves, and can also be used as approximations of the expected effect in scenarios involving either non-homogenous diffusion scenarios such as those put forth by ~\citet{fryer2007} and \citet{ballantyne_hess_tev}.

Finally, a significant uncertainty in this model pertains to the angular dependence of cosmic-ray injection from the GC point source: while a spherically symmetric distribution was found by \citet{linden_gc} to provide a compelling match to the current H.E.S.S. point source limit on the GC emission, other models are possible. We will comment on the assumption of isotropy in the cosmic ray distribution in Section~\ref{sec:conclusions}. Throughout this work, we restrict our analysis to examining photons in the energy range of 1-10 TeV, which yields several simplifications to the analysis. First, the spectrum in this region is relatively flat, following a best fit spectral index $\Gamma$~=~2.10 $\pm$ 0.04 \citep{hess_no_variability_2009}. Secondly, the PSF of both H.E.S.S. and CTA are relatively constant in this region \citep{aharonian_galactic_disk_and_angular_resolution, ctaspecs}.

%%Models of the HESS Instrument
The most up-to-date analysis of the GC  with H.E.S.S. consisted of 93h of live-time with the instrument operating in ``Wobble" mode with a an average distance of  0$^\circ$.7 from the position of Sgr A*, and producing events with a mean zenith angle of 23$^\circ$~\citep{hess_no_variability_2009}. From \citet[Fig. 13]{hess_instrument_performance_2006} we infer an effective area of 2~x~10$^9$~cm$^2$, with only negligible variation over the energy range 1-10~TeV. In this region, we adopt a point-spread function which is constant in energy and follows the functional form given in \citet{hess_instrument_performance_2006} of a two-component Gaussian where the probability of finding a photon in a radial bin d$\theta$ is given by:

\begin{equation}
\label{eq:psf}
P(\theta) = A\theta( \exp(-\frac{\theta^2}{2\sigma_1^2})+ A_{\rm rel}\exp(-\frac{\theta^2}{2\sigma_2^2}))
\end{equation}

\begin{figure}
                \plotone{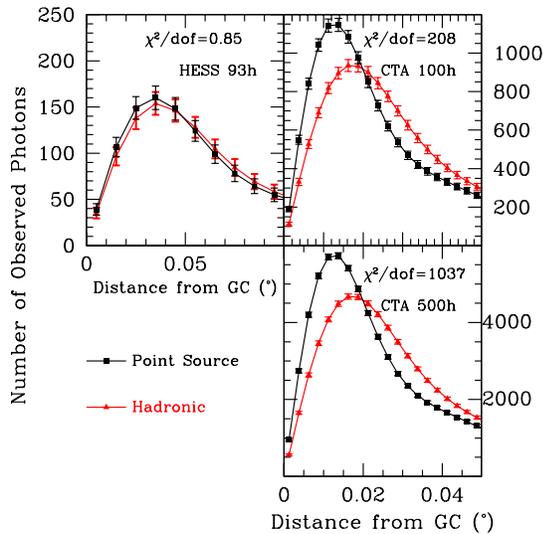}
                	\caption{ \label{fig:binneddata}Expected photon counts as a function of the distance from the GC ($^\circ$ from Sgr A*), for models where the 1-10 TeV signal is generated by photons from a point source at the position of Sgr~A* (black) or via hadronic emission from the central point source which propagates rectilinearly ($\rho$(r)~$\propto$~r$^{-2}$) and subsequently interacts with gas (red). The mean shown in each model is the average of 1000 realizations, and error bars indicate the $\sqrt{counts}$ for a the average simulation. The value of $\chi^2$/d.o.f is computed via a two-sample K-S test which does not depend on the binning used in the figure.}
\end{figure}

with $\sigma_1$~=~0.046, $\sigma_2$~=~0.12, A$_{rel}$~=~0.15 and A a normalization constant. However, in this work, these parameters are set specifically to account for the spectral characteristics of the Crab Pulsar and a 60$^\circ$ zenith angle. These parameters provide a 68\% containment angle of 0.12$^\circ$ degrees. In the case of the GC, \citet{aharonian_galactic_disk_and_angular_resolution} yields a 68\% containment angle of 0.08$^\circ$, and we thus linearly scale down the parameters $\theta_1$ and $\theta_2$ to the values $\theta_1$~=~0.031 and $\theta_2$~=~0.08 in order to obtain the correct 68\% containment angle. Finally, H.E.S.S. observations from \citep{hess_no_variability_2009} find a best fit intensity above 1~TeV of I$_{>1TeV}$~=~(1.99~$\pm$~0.09)~x~10$^{-12}$~cm$^{-2}$~s$^{-1}$. Given the calculated effective area of H.E.S.S, this implies a point source observation of 1332 photons with energy in the range 1-10~TeV. 

%%Models of the CTA instrument
In order to model the instrumental performance of CTA, we adopt best fitting parameters following the design specifications set forth in \citet{ctaspecs}, noting, however, that the ultimate design specifications for the instrument are presently unknown. Specifically, we adopt an effective area in the 1-10~TeV band of 2~x~10$^{10}$~cm$^2$, which exceeds the H.E.S.S. effective area by  an order of magnitude, and we adopt an equivalent functional form for the point-spread function as described in Eq.~(\ref{eq:psf}) for H.E.S.S., but rescale the parameters $\sigma_1$ and $\sigma_2$ such that the 68\% containment radius of the photon signal is 0.03$^\circ$. This yields $\sigma_1$~=~0.0115 and $\sigma_2$~=~0.03. While CTA contains additional improvements over current Cherenkov telescopes, especially stemming from its significantly lower energy threshold, the poorer angular resolution in the lower energy regime mitigates the effectiveness of CTA to test the morphology of the dense GC region. In this work we evaluate the performance of CTA after both 100 and 500 hours of observation, indicating both a conservative lower bound and a targeted observation time for the GC region. This yields an expected 14323 and 71613 photon counts, respectively. 

In order to simulate observations of the GC with both the H.E.S.S. and CTA instruments we employ Monte Carlo techniques to calculate the expected distribution of observed photons. We first calculate the 3D morphology for the true photon direction. In the case of hadronic emission, we calculate this by multiplying the 3D distribution of gas with both the r$^{-2}$ and r$^{-1}$ cosmic-ray densities assumed for rectilinear and diffusive transport of TeV protons. For dark matter models we assume a density distribution $\rho$(r)~=~r$^{\alpha}$ and evaluate scenarios $\alpha$=\{1.0, 1.4\}, as we note that current hydrodynamical simulations indicate the possibility that the inner dark matter density profile is adiabatically contracted (i.e. $\alpha$~$>$~1.0) \citep{blumenthal_adiabatic_contraction_1986, ryden_adiabatic_contraction_1987, gnedin_adiabatic_contraction_2004, gnedin_adiabatic_contraction_2011}. These emission profiles are integrated over the line of sight, and then photons are selected from this distribution and smeared with the PSF of each instrument. We run 1000 simulations of all models in order to achieve reasonable statistical accuracy, unless otherwise noted.

\section{Results}
\label{sec:results}

%%HESS Results
In Figure~\ref{fig:2dplot} we show $\gamma$-ray intensity maps depicting the actual emission morphology (i.e. for a machine with perfect angular resolution) (top row), the emission as observed by a machine with the angular resolution of the HESS telescope (middle row) and the emission as observed by a machine with the angular resolution of CTA (bottom row) in the case of hadronic emission in the regime rectilinear proton propagation (left column), hadronic emission in the regime of diffusive proton propagation (middle column), and point source emission. In models employing the effective area and angular resolution of the H.E.S.S. telescope, the morphological features stemming from the gas density employed in the hadronic emission scenario occur on angular scales significantly smaller than the $\sigma$~=~0.08$^\circ$ angular resolution of the telescope blurring out the angular features which distinguish the hadronic scenario. Since these angular features are themselves centered around the position of Sgr A*, differentiating between the point-source and hadronic models becomes extremely difficult.

\begin{figure}
                \plotone{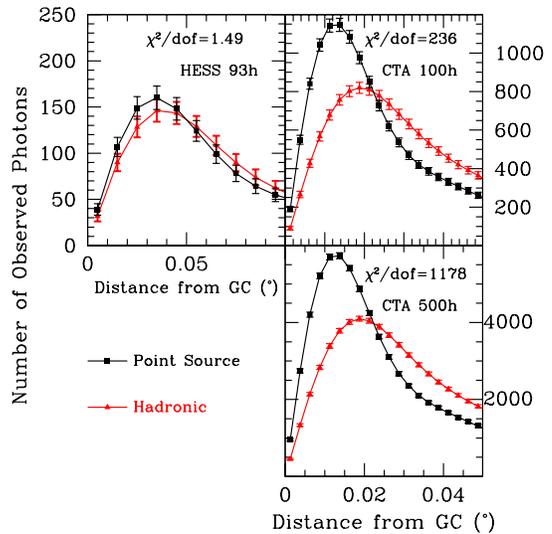}
                	\caption{ \label{fig:binneddata_diffusive} Same as Figure~\ref{fig:binneddata} but for protons which are propagating diffusively ($\rho$(r)~$\propto$~r$^{-1}$) (red)}
\end{figure}

Calculating the average cumulative-distribution function over 1000 simulations of both a GC point source and the hadronic emission scenario, we employ a K-S test and find that current H.E.S.S. results would only be able to differentiate between models with a $\chi^2$/d.o.f.~=~0.85, which falls far short of providing a minimum 2$\sigma$ level of confidence between the emission profiles.  In Figure~\ref{fig:binneddata} (top left) we provide a binned (at 0.01$^\circ$) sample of photons collected by the H.E.S.S. telescope in each scenario. In realistic observations, this result is further complicated by residuals stemming from both the Galactic plane integrated over the line of sight, as well as from contaminating cosmic-ray backgrounds, both of which should appear isotropic in the small region under consideration. While additional H.E.S.S. observations time may slightly improve these statistics, these additional backgrounds make it unlikely that the H.E.S.S. telescope will be capable of differentiating between the point source and hadronic emission scenarios.

%%CTA results
In the case of CTA, the improved angular resolution will provide a much sharper view to distinguish between a point source and hadronic models. Furthermore, the greatly increased effective area will (in our simplified model where additional backgrounds are rejected) increase the $\chi^2$ mismatch between models linearly. Using 100 hours of CTA observation, a K-S test provides a fit $\chi^2$/dof = 208, which provides more than 14$\sigma$ differentiation between models. We find that the improved angular resolution of CTA will allow for a 3$\sigma$ rejection of the poorer fitting model with only $\sim$5h of pointed observation! In Figure~\ref{fig:binneddata} (top right), we again show a binned analysis for this dataset, noting specifically the under-density of photons observed within the inner 0.01$^\circ$, which provides an independent, statistically significant indication that would be difficult  to explain with an additional diffuse or cosmic-ray background. We note that over the projected lifetime of CTA, nearly 500h of GC observation are expected, which would lead to a 32$\sigma$ differentiation between models, with a result that is plotted in Figure~\ref{fig:binneddata} (bottom right). A wealth of information on the nature of the GC source will clearly be available in this case, going well beyond simply distinguishing between a point-source emission and a hadronic model.

%%Diffusion case
In Figure~\ref{fig:binneddata_diffusive} we show the same observations as in Figure~\ref{fig:binneddata}, but under the assumption that high energy protons travel diffusively through the inner galaxy, and thus follow a density distribution $\rho$(r)~$\propto$~r$^{-1}$ instead of $\rho$(r)~$\propto$~r$^{-2}$. In this case, we find that the statistical differentiation between the Point Source and Hadronic scenarios slightly increases in all cases, due to the more diffuse nature of the energetic protons. However, the qualitative results are unchanged. Specifically, in the case of H.E.S.S observations this improvement is still insufficient to differentiate the two signals, as our K-S test obtains a fit $\chi^2$/dof = 1.49.

\begin{figure}
                \plotone{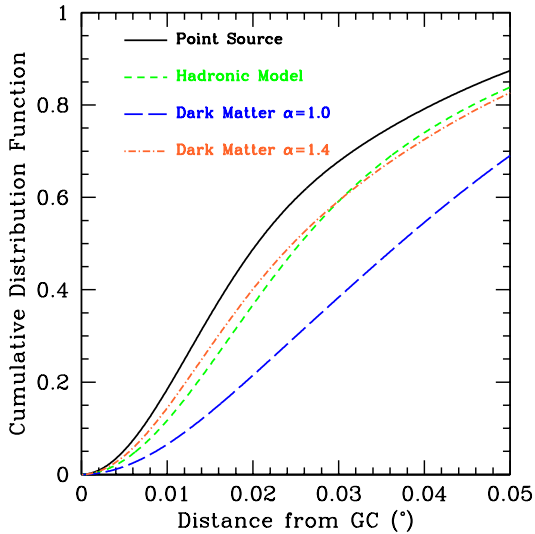}
                	\caption{ \label{fig:cumulative_distribution}Cumulative Distribution Function (CDF) of photons observed within 0.05$^\circ$ of the GC for models of the total number of photons produced within $\sim$0.07$^\circ$ of the GC. We show the case of a point source at the position of the GC (solid black), the hadronic model as described in Section~\ref{sec:models} (green short dash), dark matter following an NFW profile $\alpha$~=~1.0 (blue long dash) and dark matter following a steeper profile with $\alpha$~=~1.4 (orange dot-dashed).}
\end{figure}

%%Dark Matter can confuse this source
In realistic models, additional emission sources must also be considered, including an isotropic cosmic-ray background, a line of sight background through the Galactic plane\footnote{We note that this is approximately isotropic for the very small angular regions considered here.}, and unresolved sources in the region surrounding the GC -- all of which will contribute additional uncertainties to the differentiation of the point source and hadronic models. One particularly interesting background could stem from the annihilation of dark matter particles in the GC region. The morphology of the dark matter annihilation is partially constrained to be spherically symmetric with a density distribution which follows a form $\rho$(r)~$\propto$~r$^{-\alpha}$. While a standard value, $\alpha$~=~1.0 is employed in the standard NFW dark matter model \citep{nfw}, the dark matter profile is highly uncertain in the GC region, and the gravitational effect from baryons in the GC may significantly steepen the dark matter distribution~\citep{blumenthal_adiabatic_contraction_1986, ryden_adiabatic_contraction_1987, gnedin_adiabatic_contraction_2004, gnedin_adiabatic_contraction_2011}. This makes it potentially difficult to differentiate between dark matter models for TeV emission from the GC, and the possible combination of emission from both point source and hadronic sources. 

\begin{figure}
                \plotone{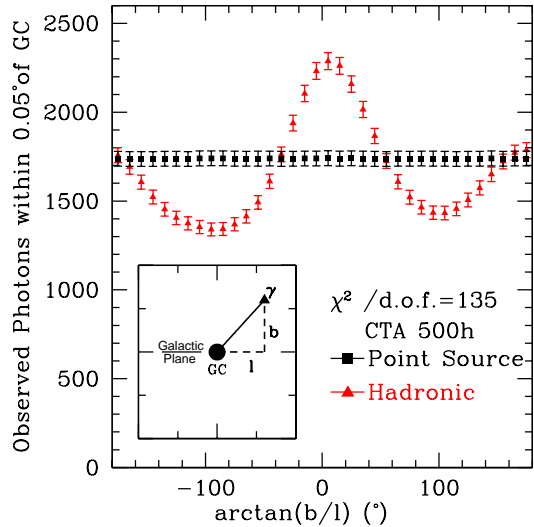}
                	\caption{ \label{fig:angulardata}Expected photon counts as a function of the azimuthal angle $\phi=\arctan(b/l)$ (shown in the inset),  for both the point-source (black squares) and the hadronic emission scenario for protons which propagate rectilinearly (red triangles) for photons observed within 0.05$^\circ$ of the GC. In the case of a point source, photons are symmetrically distributed around the position of the GC, while the majority of emission due to hadronic injection follows the distribution of gas, which is aligned more closely with the Galactic plane.}
\end{figure}

In Figure~\ref{fig:cumulative_distribution} we plot a projection for the cumulative distribution function of photons observed with 500h at CTA as a function of the angular distance from the GC, for models of point source emission, hadronic emission, dark matter annihilation with a density profile $\alpha$~=~1.0 and with a density profile $\alpha$~=~1.4. We note that the point source model contains the fastest rising CDF possible, with a morphology uniquely determined by the instrumental PSF. Any combination of emission from a point source and the hadronic model must produce a CDF which lies between the individual models, and the relative contribution of each source class can be accurately (to within $\sim$10\% errors after 500h of observation) determined by examining the CDF observed by CTA. However, a small contribution from dark matter annihilation following an index $\alpha$~=~1.0 along with a dominant contribution from the point-source, may be misinterpreted as emission stemming from important contributions of both the point-source and hadronic models. Moreover, models where the emission is entirely dominated by dark matter which is highly peaked towards the GC (such as $\alpha$~=~1.4), may also be misinterpreted as some linear combination of hadronic and point source contributions. This uncertainty is a standard result from an attempt to identify three unknown intensities using only one constraint parameter.

\begin{figure}
                \plotone{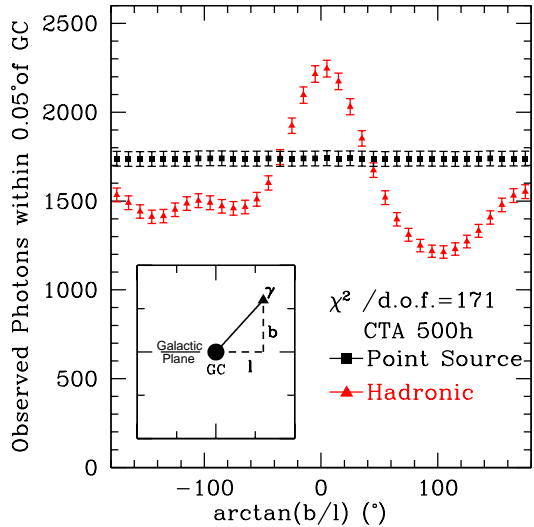}
                	\caption{ \label{fig:angulardata_diffusive} Same as Figure~\ref{fig:angulardata} but for protons which are propagating diffusively ($\rho$(r)~$\propto$~r$^{-1}$) (red)}
\end{figure}

%%CTA results in detecting non-spherical behavior
A separate measurement is therefore necessary in order to constrain the relative contributions from all three source classes. An obvious choice is to model the angular distribution of photons around the GC, noting that both the point source and dark matter models are spherically symmetric. In Figure~\ref{fig:angulardata} we plot the expected azimuthal angular distribution for both the point source and hadronic models, counting the number of photons from a given angle $\phi=\arctan(b/l)$, i.e. the angle formed between the Galactic plane and the direction joining the GC and the photon location in the sky (see inset). We restrict the counts to photons within 0.05$^\circ$ of the GC, where contributions from the H.E.S.S. point source are believed to be largely dominant. While a (spherically symmetric) point source provides a flat distribution in the incoming photon angle (as expected), contributions from the hadronic scenario deviate significantly, and are primarily aligned with the Galactic plane, due to contributions from both the circum-nuclear ring, as well as a contribution at angles of near 0$^\circ$ from hadronic interactions within the molecular cloud M-0.02-0.07. We note that with 500h of observation, an evaluation of the nature of the central TeV source can be made with more than 11$\sigma$ confidence - without any reference to the radial distribution shown in Figure~\ref{fig:binneddata}. Most importantly, this implies that in cases where the point-source, hadronic sources, and dark matter annihilation all contribute non-negligibly to the total TeV galactic center source, we can determine the relative contribution stemming from hadronic emission to within 20\% with more than 2$\sigma$ confidence. In Figure~\ref{fig:angulardata_diffusive} we show the same analysis as in Figure~\ref{fig:angulardata}, under the assumption of diffusively propagating protons and find the results to be qualitatively unchanged, with a $\chi^2$/dof which improves by about 25\%. 

\section{Discussion and Conclusions}
\label{sec:conclusions}

We have shown that while current TeV instruments are incapable of distinguishing between point-source and hadronic emission models for the GC, the upcoming CTA will definitively differentiate between the signals within its first hours of observation at the GC, and will determine the relative importance of each contributing source class at the 10\% level over the course of its lifetime (see Fig.~\ref{fig:binneddata} and related discussion). These source classes can be distinguished based on either their radial or angular morphology, yielding two independent handles for the determination of the $\gamma$-ray source. The angular information is particularly important, as it allows the hadronic scenario to be easily separated from {\em any} combination of a GC point source and models of dark matter annihilation.

While in this work, we have considered only $\gamma$-rays with an energy exceeding 1~TeV in our quantitative evaluation of the ability of CTA to distinguish between the hadronic and point-source signals, we note that the discrimination power of CTA could be further extended depending on the specific instrumental point spread function which the instrument attains at lower energies. The continuation, or even softening of the E$^{-2.0}$ $\gamma$-ray spectrum to energies as low as 1~GeV has been demonstrated by the combination of HESS \citep{hess_no_variability_2009} and Fermi-LAT~\citep{vitale_galactic_center, hooper_goodenough_gc, hooper_linden_gc} results., This has the potential to create a much stronger statistical test, so long as the angular resolution of CTA remains on the same scale as the size of the circum-nuclear ring. Moreover, the differentiation between the Hadronic and point-source models improves in the diffusive regime, which is thought to control the propagation of the protons which create the $\gamma$-ray signal below $\sim$100~GeV~\citep{chernyakova, linden_gc}. 

We further note that while this study quantitatively examines only one specific scenario for hadronic diffusion through the GC region where the diffusion constant was fine tuned in order to match the softening of the $\gamma$-ray spectrum at energies $\sim$100~GeV~\citep{chernyakova}, the model remains relevant for other physical models which have been proposed to control the propagation of high energy protons in the galactic center region. Specifically, \citet{fryer2007} employs a model which allows rectilinear propagation out to approximately 1~pc, after which the shocked winds produce a proton density which remains relatively flat between 1-3~pc, corresponding with the region dominated by the circum-nuclear ring. The comparison of our results in the case of both rectilinear ($\rho$(r)~$\propto$~r$^{-2}$) and diffusive ($\rho$(r)~$\propto$~r$^{-1}$) propagation show that the overall change in the pion production morphology is relatively unchanged by this factor of 3 change in the proton density at the outer edge of the circum-nuclear ring. In fact, from our results, we would expect a marginal improvement in the differentiation between the model of \citep{fryer2007} and that of point-source $\gamma$-ray production. The same is also true in the models of \citep{ballantyne_hess_tev}, where the diffusion constant varies inversely with the local gas density, producing a density of cosmic-rays which increases greatly in the region of the circum-nuclear ring.

There is however, one important caveat to the scenario employed here. The morphology of observed high energy gamma-rays from the Hadronic scenario may become significantly less extended in cases where high energy protons are not isotropized by the galactic medium through which they are propagating. In this case, $\gamma$-rays observed at the solar position will preferentially stem from protons which were themselves originally pointed towards the solar position, due to the relativistic beaming of pions produced in p-p scattering. This complication may become important in scenarios where we have assumed rectilinear proton propagation, which in the case of a simple, homogeneous diffusion constant, would imply that protons do not become completely isotropized before leaving the diffusion zone. In this scenario, we find that for diffusion constants exceeding 9.3~$\times$~10$^{29}$~cm$^2$~s$^{-1}$, the random walk approximation for particle diffusion predicts that the average proton will not undergo a change in direction before leaving the 10~pc diffusion region. In the studies of \citet{chernyakova} and \citet{linden_gc} this corresponds to a proton energy of only 2.2~TeV. It is these protons which dominate the 1~TeV signal, since the E$^{-2}$ proton injection signal implies that the most important contributor to $\gamma$-rays of a given energy are protons of a very similar energy. This could mean that some fraction of protons which are undergoing p-p scattering in our simulations are not isotropically distributed.

However, the random-walk approximation for proton diffusion is notably poor in this region, and all protons will undergo some scattering off of the magnetic field inhomogenities in the diffusion region. The exact degree of proton anisotropy will depend sensitively on the specifics of this diffusion scenario, and specifically on any inhomogenities in the magnetic field structure, which are both invariably present, and difficult to directly determine. Lastly, we note that even if the vast majority of the signal is anisotropic (thus appearing very similar to the point source calculation), even a small isotropic component could be distinguished from the point source calculation, given the incredible $\chi^2$ detection of the Hadronic model by the CTA.

We note that this same process could also place constraints on the dark matter annihilation cross-section -  using the observed angular and radial profiles to determine the maximum contribution of photons following a morphology consistent with dark matter annihilation. However, due to the small angular region considered, as well as the extremely bright point-source emission observed by H.E.S.S., these constraints are not competitive with those determined, for example, by H.E.S.S. observations of the regions directly above and below the Galactic plane~\citep{hess_gc_2011}. While the GC remains an extremely interesting region for setting dark matter constraints with CTA, the best limits will likely continue to be set by analysis of regions directly off of the Galactic plane. This scenario may become interesting, however, in cases where the dark matter density in the GC region is found to be {\em highly} adiabatically contracted (e.g. $\alpha$~~$\gtrsim$~1.7), leading to extremely enhanced fluxes within the inner pc of the Galaxy.

Pinpointing the nature of the GC source will allow us to more effectively search for a dark matter signal in this region, as the differentiation between the point source and hadronic models would allow for an extrapolation of their expected emission profiles into the regions where strong dark matter limits can be set. Additionally, the extrapolation of the observed CTA emission will greatly refine Fermi-LAT models of the morphology of the central point source, as shown in~\citet{linden_gc}. Thus TeV $\gamma$-ray observations will be critical to validate or constrain particle dark matter models that could explain other observations tentatively indicating signals from dark matter annihilation at lower energies (e.g \citet{dan_10_GeV}). In this scenario, the role of CTA will be highly complementary to an extended Fermi-LAT campaign, especially in accurately subtracting a diffuse signal emission from the bright, and possibly extended, central source.

\acknowledgments
We are grateful to Katia Ferriere and David Williams for helpful comments. This work is partly supported by NASA grant NNX11AQ10G. SP acknowledges support from an Outstanding Junior Investigator Award from the Department of Energy, and from DoE grant DE-FG02-04ER41286. \\ %\newpage

\bibliography{cta} 

\begin{thebibliography}{33}
\expandafter\ifx\csname natexlab\endcsname\relax\def\natexlab#1{#1}\fi

\bibitem[{{Abazajian}(2011)}]{abazajian_msp_gc}
{Abazajian}, K.~N. 2011, JCAP, 3, 10

\bibitem[{{Abramowski} {et~al.}(2011){Abramowski}, {Acero}, {Aharonian},
  {Akhperjanian}, {Anton}, {Barnacka}, {Barres de Almeida}, {Bazer-Bachi}, \&
  {et al.}}]{hess_gc_2011}
{Abramowski}, A., {Acero}, F., {Aharonian}, F., {Akhperjanian}, A.~G., {Anton},
  G., {Barnacka}, A., {Barres de Almeida}, U., {Bazer-Bachi}, A.~R., \& {et
  al.} 2011, Physical Review Letters, 106, 161301

\bibitem[{{Acero} {et~al.}(2010){Acero}, {Aharonian}, {Akhperjanian}, {Anton},
  {Barres de Almeida}, {Bazer-Bachi}, {Becherini}, {Behera}, \& {et
  al.}}]{acera_hess_gc_position_2010}
{Acero}, F., {Aharonian}, F., {Akhperjanian}, A.~G., {Anton}, G., {Barres de
  Almeida}, U., {Bazer-Bachi}, A.~R., {Becherini}, Y., {Behera}, B., \& {et
  al.} 2010, \mnras, 402, 1877

\bibitem[{{Aharonian} {et~al.}(2009){Aharonian}, {Akhperjanian}, {Anton},
  {Barres de Almeida}, {Bazer-Bachi}, {Becherini}, {Behera}, {Bernl{\"o}hr}, \&
  {et al.}}]{hess_no_variability_2009}
{Aharonian}, F., {Akhperjanian}, A.~G., {Anton}, G., {Barres de Almeida}, U.,
  {Bazer-Bachi}, A.~R., {Becherini}, Y., {Behera}, B., {Bernl{\"o}hr}, K., \&
  {et al.} 2009, \aap, 503, 817

\bibitem[{{Aharonian} {et~al.}(2004){Aharonian}, {Akhperjanian}, {Aye},
  {Bazer-Bachi}, {Beilicke}, {Benbow}, {Berge}, {Berghaus}, \& {et
  al.}}]{2004A&A...425L..13A}
{Aharonian}, F., {Akhperjanian}, A.~G., {Aye}, K.-M., {Bazer-Bachi}, A.~R.,
  {Beilicke}, M., {Benbow}, W., {Berge}, D., {Berghaus}, P., \& {et al.} 2004,
  \aap, 425, L13

\bibitem[{{Aharonian} {et~al.}(2008){Aharonian}, {Akhperjanian}, {Barres de
  Almeida}, {Bazer-Bachi}, {Becherini}, {Behera}, {Benbow}, {Bernl{\"o}hr}, \&
  {et al.}}]{hess_no_variability_xray}
{Aharonian}, F., {Akhperjanian}, A.~G., {Barres de Almeida}, U., {Bazer-Bachi},
  A.~R., {Becherini}, Y., {Behera}, B., {Benbow}, W., {Bernl{\"o}hr}, K., \&
  {et al.} 2008, \aap, 492, L25

\bibitem[{{Aharonian} {et~al.}(2006{\natexlab{a}}){Aharonian}, {Akhperjanian},
  {Bazer-Bachi}, {Beilicke}, {Benbow}, {Berge}, {Bernl{\"o}hr}, {Boisson}, \&
  {et al.}}]{hess_galactic_center_dm_2006}
{Aharonian}, F., {Akhperjanian}, A.~G., {Bazer-Bachi}, A.~R., {Beilicke}, M.,
  {Benbow}, W., {Berge}, D., {Bernl{\"o}hr}, K., {Boisson}, C., \& {et al.}
  2006{\natexlab{a}}, Physical Review Letters, 97, 221102

\bibitem[{{Aharonian} {et~al.}(2006{\natexlab{b}}){Aharonian}, {Akhperjanian},
  {Bazer-Bachi}, {Beilicke}, {Benbow}, {Berge}, {Bernl{\"o}hr}, {Boisson}, \&
  {et al.}}]{hess_instrument_performance_2006}
---. 2006{\natexlab{b}}, \aap, 457, 899

\bibitem[{{Aharonian} {et~al.}(2006{\natexlab{c}}){Aharonian}, {Akhperjanian},
  {Bazer-Bachi}, {Beilicke}, {Benbow}, {Berge}, {Bernl{\"o}hr}, {Boisson}, \&
  {et al.}}]{aharonian_galactic_disk_and_angular_resolution}
---. 2006{\natexlab{c}}, \apj, 636, 777

\bibitem[{{Aharonian} \& {Neronov}(2005)}]{aharonian_nernov_hadronic_gc_2005}
{Aharonian}, F., \& {Neronov}, A. 2005, \apss, 300, 255

\bibitem[{{Atwood} {et~al.}(2009){Atwood}, {Abdo}, {Ackermann}, {Althouse},
  {Anderson}, {Axelsson}, {Baldini}, {Ballet}, {Band}, {Barbiellini}, \&
  et~al.}]{2009ApJ...697.1071A}
{Atwood}, W.~B., {Abdo}, A.~A., {Ackermann}, M., {Althouse}, W., {Anderson},
  B., {Axelsson}, M., {Baldini}, L., {Ballet}, J., {Band}, D.~L.,
  {Barbiellini}, G., \& et~al. 2009, \apj, 697, 1071

\bibitem[{{Ballantyne} {et~al.}(2007){Ballantyne}, {Melia}, {Liu}, \&
  {Crocker}}]{ballantyne_hess_tev}
{Ballantyne}, D.~R., {Melia}, F., {Liu}, S., \& {Crocker}, R.~M. 2007, \apjl,
  657, L13

\bibitem[{{Becklin} {et~al.}(1982){Becklin}, {Gatley}, \&
  {Werner}}]{circumnuclear_ring_1982}
{Becklin}, E.~E., {Gatley}, I., \& {Werner}, M.~W. 1982, \apj, 258, 135

\bibitem[{{Blumenthal} {et~al.}(1986){Blumenthal}, {Faber}, {Flores}, \&
  {Primack}}]{blumenthal_adiabatic_contraction_1986}
{Blumenthal}, G.~R., {Faber}, S.~M., {Flores}, R., \& {Primack}, J.~R. 1986,
  \apj, 301, 27

\bibitem[{{Bradford} {et~al.}(2005){Bradford}, {Stacey}, {Nikola}, {Bolatto},
  {Jackson}, {Savage}, \& {Davidson}}]{2005ApJ...623..866B}
{Bradford}, C.~M., {Stacey}, G.~J., {Nikola}, T., {Bolatto}, A.~D., {Jackson},
  J.~M., {Savage}, M.~L., \& {Davidson}, J.~A. 2005, \apj, 623, 866

\bibitem[{{Chernyakova} {et~al.}(2011){Chernyakova}, {Malyshev}, {Aharonian},
  {Crocker}, \& {Jones}}]{chernyakova}
{Chernyakova}, M., {Malyshev}, D., {Aharonian}, F.~A., {Crocker}, R.~M., \&
  {Jones}, D.~I. 2011, \apj, 726, 60

\bibitem[{{CTA Consortium}(2011)}]{ctaspecs}
{CTA Consortium}, T. 2011, ArXiv e-prints

\bibitem[{{Ferri{\`e}re}(2012)}]{ferriere_2012_gas_density}
{Ferri{\`e}re}, K. 2012, \aap, 540, A50

\bibitem[{{Fryer} {et~al.}(2007){Fryer}, {Liu}, {Rockefeller}, {Hungerford}, \&
  {Belanger}}]{fryer2007}
{Fryer}, C.~L., {Liu}, S., {Rockefeller}, G., {Hungerford}, A., \& {Belanger},
  G. 2007, \apj, 659, 389

\bibitem[{{Gnedin} {et~al.}(2011){Gnedin}, {Ceverino}, {Gnedin}, {Klypin},
  {Kravtsov}, {Levine}, {Nagai}, \&
  {Yepes}}]{gnedin_adiabatic_contraction_2011}
{Gnedin}, O.~Y., {Ceverino}, D., {Gnedin}, N.~Y., {Klypin}, A.~A., {Kravtsov},
  A.~V., {Levine}, R., {Nagai}, D., \& {Yepes}, G. 2011, ArXiv e-prints

\bibitem[{{Gnedin} {et~al.}(2004){Gnedin}, {Kravtsov}, {Klypin}, \&
  {Nagai}}]{gnedin_adiabatic_contraction_2004}
{Gnedin}, O.~Y., {Kravtsov}, A.~V., {Klypin}, A.~A., \& {Nagai}, D. 2004, \apj,
  616, 16

\bibitem[{{Hooper}(2012)}]{dan_10_GeV}
{Hooper}, D. 2012, ArXiv e-prints

\bibitem[{{Hooper} {et~al.}(2004){Hooper}, {de la Calle Perez}, {Silk},
  {Ferrer}, \& {Sarkar}}]{hooper_tev_dm_2004}
{Hooper}, D., {de la Calle Perez}, I., {Silk}, J., {Ferrer}, F., \& {Sarkar},
  S. 2004, JCAP, 9, 2

\bibitem[{{Hooper} \& {Goodenough}(2011)}]{hooper_goodenough_gc}
{Hooper}, D., \& {Goodenough}, L. 2011, Physics Letters B, 697, 412

\bibitem[{{Hooper} {et~al.}(2012){Hooper}, {Kelso}, \&
  {Queiroz}}]{hooper_kelso_gc}
{Hooper}, D., {Kelso}, C., \& {Queiroz}, F.~S. 2012, ArXiv e-prints

\bibitem[{{Hooper} \& {Linden}(2011)}]{hooper_linden_gc}
{Hooper}, D., \& {Linden}, T. 2011, \prd, 84, 123005

\bibitem[{{Linden} {et~al.}(2012){Linden}, {Lovegrove}, \&
  {Profumo}}]{linden_gc}
{Linden}, T., {Lovegrove}, E., \& {Profumo}, S. 2012, \apj, 753, 41

\bibitem[{{Liu} {et~al.}(2006{\natexlab{a}}){Liu}, {Melia}, {Petrosian}, \&
  {Fatuzzo}}]{liu_hadronic_gc_2006a}
{Liu}, S., {Melia}, F., {Petrosian}, V., \& {Fatuzzo}, M. 2006{\natexlab{a}},
  \apj, 647, 1099

\bibitem[{{Liu} {et~al.}(2006{\natexlab{b}}){Liu}, {Petrosian}, {Melia}, \&
  {Fryer}}]{liu_hadronic_gc_2006b}
{Liu}, S., {Petrosian}, V., {Melia}, F., \& {Fryer}, C.~L. 2006{\natexlab{b}},
  \apj, 648, 1020

\bibitem[{Navarro {et~al.}(1997)Navarro, Frenk, \& White}]{nfw}
Navarro, J.~F., Frenk, C.~S., \& White, S.~D. 1997, Astrophys.J., 490, 493

\bibitem[{{Profumo}(2005)}]{profumo_tev_dm_2005}
{Profumo}, S. 2005, \prd, 72, 103521

\bibitem[{{Ryden} \& {Gunn}(1987)}]{ryden_adiabatic_contraction_1987}
{Ryden}, B.~S., \& {Gunn}, J.~E. 1987, \apj, 318, 15

\bibitem[{{Vitale} {et~al.}(2009){Vitale}, {Morselli}, \& {for the Fermi/LAT
  Collaboration}}]{vitale_galactic_center}
{Vitale}, V., {Morselli}, A., \& {for the Fermi/LAT Collaboration}. 2009, ArXiv
  e-prints

\end{thebibliography}

\end{document}